\documentclass[11pt]{article}
    \usepackage[OT2,T1]{fontenc}
    
    \usepackage[dvips]{graphicx}
    \usepackage{amsmath}
    \usepackage{amssymb}
    \usepackage{setspace}
    \usepackage{txfonts}
    \usepackage{pxfonts}

\begin{document}
\title {Special Relativity\\
as\\
Classical kinematics  of a particle with the upper bound on its speed\\
\vspace{4.5pt}
Part II\\
\vspace{4.5pt}
The general Lorentz transformation and\\
\vspace{-1.5pt}
The generalized velocity composition theorem}
\bigskip
\author{Alex Granik\thanks{University of the Pacific, Physics Department, Stockton, CA 95211, USA. Email: agranik@pacific.edu.} }
\date{}
\maketitle
{}
\bigskip

\begin{abstract}

The kinematics of a particle with the upper bound on the particle's speed (a modification of classical kinematics where such a restriction is absent) has been developed in \cite{1}. It was based solely on classical  mechanics without employing any concepts , associated with the time dilatation or/and length contraction. It yielded the 1-D Lorentz transformation (LT),  free of inconsistencies (inherent in the canonical derivation and interpretations of the LT). Here we apply the same approach  to derive the LT  for the 3-dimensional motion of a particle and the attendant law of velocity composition. As a result, the infinite set of four-parameter transformations is obtained. The requirement of linearity of these transformations selects out of this set the two-parameter subset . The values of the remaining two parameters ,	dictated by physics of the motion, is explicitly determined , yielding  the canonical form of  the 3-dimensional LT. The generalized law of velocity composition and the attendant invariant ( not postulated $apriori$) of the motion are derived, As in the one-dimensional case, present  derivation, as a whole, does not have any need in introducing the concepts of the time dilatation and length contraction, and is based on the classical concepts of time and space.
\end{abstract}
 \thispagestyle{empty}
\section{Introduction}

\hspace{20pt}We consider the physical set-up analogous the one used in \cite{1} for describing the modified kinematics of particles with the upper bound on their speed, this time generalizing it to 3-dimensional space. To this end we consider three particles {\bf{A, B}},  and  {\bf{D}}  involved in a compound motion: Particle {\bf{B}}  moves with a constant velocity {\bf{v}}(a vector) with respect to particle {\bf{A}}, and particle {\bf{D}}  moves with respect to particle {\bf{B}}
with velocity {\bf{u'}}, not necessarily constant. If these three particle would be stationary, then the familiar identity of the stationary geometry holds true
	\begin{equation}
\label{1}
{\bf{AD=AB+BD}}
\end{equation}
None of the distances changes with time. Therefore their lengths are fixed and independent of time.

 However , when particles {\bf{B}} and {\bf{D}} move,  this identity  holds true only if  all three distances are defined at the same time, which is impossible, since their locations continuously change with time. As a result, the identity (\ref{1}) is violated, and now
\begin{equation}
\label{2}
{\bf{AD \neq AB+BD}}
\end{equation}
	This means , in particular, that if positional vectors {\bf{AB}} and {\bf{BD}} are defined at some prior moment of time $t'$ , then vector{\bf{AD}}  is defined at some later moment of time, $t$. The physical foundation (which is distinctly not the time dilatation) of the  difference between $t $ and $t'$  will be given in the forthcoming paper \cite{2}. Here we only mention that  A.Poincare in his 1898 paper \cite{3} already pointed out the existence of such a  difference (although in  rather abstract  terms) ,\

{" \it{an eclipse of the moon $...$ suppose that this phenomenon is perceived simultaneously from all points of the earth. That is not altogether true, since the propagation of light is not instantaneous; if absolute exactitude were desired, there would be a correction to make according to a complicated rule.}"}\\
He definitely did not know what was  this rule (or formulas describing it), but clearly indicated that it appears due to the time  retardation ( not the time dilatation) caused by the finiteness of the speed of light.

We introduce the following notations.
\begin{align*}
{\bf{AB'}}(t')&={\bf{v}}t', \hspace{3.5pt}{\bf{AB}}(t)={\bf{v}}t ,\\
{\bf{BD'}}(t')&={\bf{r'}}(t'),\hspace{7.5pt}{\bf{BD}}(t)={\bf{r}}(t),\\
{\bf{AD'}}(t')&={\bf{r}}(t'),\hspace{4.5pt}{\bf{AD}}(t)={\bf{r}}(t).
\end{align*}	
For the following we represent the positional vectors {\bf{r}}(t) and  {\bf{r}}(t')  as the geometric sum of their components along {\bf{v}} and the components in the direction perpendicular to {\bf{v}}.
 \begin{align}
 \label{3}
  {\bf{r}}(t')={\bf{r'}}_{{\bf{v}}}(t')+{\bf{r'}}_{\bot}(t'),\hspace{15pt}{\bf{r}}(t)={\bf{r}}_{{\bf{v}}}(t)+{\bf{r}}_{\bot}(t)
    \end{align}
 This representation is justified as long, as every vector is considered  at the same moment of time. Otherwise, the conventional geometric addition of  $2$ vectors , Eq.(1) is not valid anymore.
In (\ref{3})
\begin{align}
\label{4}
{\bf{r'}}_{\bf{v}}(t')={\bf{v}}\frac{{\bf{r'}}(t')\bullet{\bf{v}}}{v^2},\hspace{15pt}{\bf{r}}_{\bf{v}}(t)=
{\bf{v}}\frac{{\bf{r}}(t)\bullet{\bf{v}}}{v^2}
\end{align}are the components of  {\bf{ r'}}(t') and {\bf{ r}}(t) along vector {\bf{v}} , and
\begin{align}
\label{5}
\mathbf{r}'_ \bot  (t') \equiv \frac{{\mathbf{v} \times (\mathbf{r}'(t') \times \mathbf{v})}}
{{v^2 }},\quad \mathbf{r}_ \bot  (t) \equiv \frac{{\mathbf{v} \times (\mathbf{r}(t) \times \mathbf{v})}}
{{v^2 }}
\end{align}
are their components perpendicular to {\bf{v}}.
\section{Derivation of Lorentz transformation}

\hspace{20pt}As follows from (2) the component   of  the positional  vector of point $D$  at the moment of time $t$ differs from its  value  at the moment of time  $t'$  by a numerical factor which we denote by $\mu_1$ .  By introducing this factor , we transform the inequality (2) into an equation. Thus, the geometrical law of addition of  2 vectors, Eq.(1), has to be replaced by a more general vector composition law (cf.[1]):
\begin{align}
\label{6}
\mathbf{r}_\mathbf{v} (t)& = \mu _1 [\mathbf{r}'_\mathbf{v} (t') + \mathbf{v}t'] = \mu _1 \left( {\mathbf{v}\frac{{\mathbf{r}'(t')\bullet \mathbf{v}}}
{{v^2 }} + \mathbf{v}t'} \right)
\end{align}
Inversely (cf. [1]) with ${\bf{AD}}={\bf{r}}(t)$  and  ${\bf{AB}}={\bf{v}}t$   taken at the same moment of time  $t$
\begin{align}
\label{7}
\mathbf{r}'_\mathbf{v} (t') = \mu _2 [\mathbf{r}_\mathbf{v} (t) - \mathbf{v}t] = \mu _2 \left( {\mathbf{v}\frac{{\mathbf{r}(t) \bullet\mathbf{v}}}
{{v^2 }} - \mathbf{v}t} \right)
\end{align}
On the other hand, generally speaking,
\begin{align}
\label{8}
\mathbf{r}_ \bot  (t) = \mu _3 \mathbf{r}_ \bot  '(t') + \mu _4 \frac{{\mathbf{r}_ \bot  '(t') \times \mathbf{v}}}
{v}
\end{align}
where  $\mu_3$  and $\mu_4$   are scalar quantities. Thus we have to find the scalar parameters $\mu_1$ ,$\mu_2$ ,$\mu_3$  , $\mu_4$ , which appear (cf.with the  geometric addition of stationary vectors  $\mu_2=\mu_3=\mu_4=0$  , and  $\mu_1=1$ )
 in equations (\ref{6}) - (\ref{8}), because now vectors ${\bf{r}}(t)$ and  ${\bf{r'}}(t')$  are considered at different moments of time $t$ and $t'$ .

To find these parameters we use the existence of the upper bound on a speed of a particle ( not present in conventional classical mechanics) and the linearity of the resulting transformation from ${\bf{r}}(t)$  to  ${\bf{r'}}(t')$ .
In what follows we write ${\bf{r}}(t)={\bf{r}}$  and ${\bf{r'}}(t)={\bf{r'}}$ , to simplify the notations. By using (\ref{4})- (\ref{6}) and (\ref{8}) in  the second equation of  (\ref{3}) we get
\begin{align}
\label{9}
\mathbf{r} = \mu _3 \mathbf{r}' + \mathbf{v}[(\mu _1  - \mu _3 )\frac{{\mathbf{r}'\bullet\mathbf{v}}}
{{v^2 }} + \mu _1 t'] + \mu _4 \frac{{\mathbf{r}' \times \mathbf{v}}}
{v}
\end{align}
With the help of (6) and (7) we find the respective times $t $  and $t'$ . From (6) follows
	 	\begin{align}
\label{10}
 \mathbf{r}'_\mathbf{v}  = \frac{{\mathbf{r}_\mathbf{v} }}
{{\mu _1 }} - \,\mathbf{v}t'
\end{align}
Replacing in (\ref{10}) the left hand side by its expression from (\ref{7}) and constructing the dot product of both parts of the resulting equation with ${\bf{v}}$, we get
\begin{align*}
	\mu _2 \left( {vr\cos \Theta  - v^2 t} \right) = \frac{{vr\cos \Theta }}
{{\mu _1 }} - v^2 t'
\end{align*}
yielding
\begin{align}
\label{11}
	t' = \mu _2 t + (1 - \mu _1 \mu _2 )\frac{{r\cos \Theta }}
{{\mu _1 v}}
\end{align}
Here $\Theta$  is the angle between vectors ${\bf{r}}$ and ${\bf{v}}$
\begin{align*}
r\cos \Theta  \equiv \frac{{\mathbf{rv}}}
{v} = \frac{{(\mathbf{r}_\mathbf{v}  + \mathbf{r}_ \bot  )\mathbf{v}}}
{v} \equiv \frac{{\mathbf{r}_\mathbf{v} \mathbf{v}}}
{v} = \mu _1 (\frac{{\mathbf{r}'\bullet\mathbf{v}}}
{v} + vt')
\end{align*}
where we use (6). Inserting this relation in (11) we find $t$
\begin{align}
\label{12}
t = \mu _1 t' + \frac{{\mu _1 \mu _2  - 1}}
{{\mu _2 }}\frac{{\mathbf{r}'\bullet\mathbf{v}}}
{{v^2 }}
\end{align}
Since determination of  ${\bf{r'}}$  requires somewhat lengthy calculations, we provide them in Appendix $A.$ The result is
\begin{align}
\label{13}
{\bf r}' = \frac{{\mu _3 }}{{\mu _3 ^2  + \mu _4 ^2 }}{\bf r} + {\bf v}[(\mu _2  - \frac{{\mu _3 }}{{\mu _3 ^2  + \mu _4 ^2 }})\frac{{r\cos \Theta }}{v} - \mu _2 t] - \frac{{\mu _4 }}{{\mu _4 ^2  + \mu _3 ^2 }}{\bf r} \times {\bf v}
\end{align}
where we used the identity
\begin{align*}
\frac{{{\bf a} \times ({\bf b} \times {\bf a})}}{{a^2 }} \equiv {\bf b} - {\bf a}\frac{{{\bf b\bullet a}}}{{a^2 }}
\end{align*}

In what follows we will use the assumption (see [1]) of  the existence of the universal upper limit (let us denote it by $c$) on the speed of the particles  $u^2 _{\max }  = u'^2 _{\max }  \equiv c^2 $ , where
\begin{align}
\label{14}
c^2  = u^2 _{\max }  = \left( {\frac{{dr }}{{dt}}} \right)_{\max }^2 ,\;c^2  = u'^2 _{\max }  = \left( {\frac{{dr' }}{{dt' }}} \right)_{\max }^2
\end{align}
Since $c=const$  we can use in (14) instead of the derivatives the ratios of the respective variables $r/t$ and $r'/t'$. Squaring both sides of (11) we get
\begin{align}
\label{15}
 (t')^2 & = [ {\mu _2 t + (1 - \mu _1 \mu _2 )\frac{{r\cos \Theta }}{{\mu _1 v}}} ]^2 \nonumber =\\
  &= \mu _2 ^2 t^2  + \left( {\frac{{1 - \mu _1 \mu _2 }}{{\mu _1 }}} \right)^2 \left( {\frac{{r\cos \Theta }}{v}} \right)^2 \nonumber + \\
 & + 2\mu _2 t\frac{{1 - \mu _1 \mu _2 }}{{\mu _1 }}\frac{{r\cos \theta }}{v}\Theta
 \end{align}
By introducing  the auxiliary notations  $\mu_3/(\mu_3^2+\mu_4^2)\equiv\alpha, \mu_4/(\mu_3^2+\mu_4^2)\equiv\beta$ and squaring (13), we obtain after straightforward calculations
\begin{align}
\label{16}
 (r')^2  &= {\alpha {\bf r} + {\bf v}[(\mu _2  - \alpha )\frac{r\cos \Theta }{v} - \mu _2 t] - \beta \frac{{\bf r} \times {\bf v}}{v}} ^2  =\nonumber  \\
 & = \mu _2 ^2 [r\cos \Theta  - vt]^2  + r^2 (\alpha ^2  + \beta ^2 )\sin ^2 \Theta
 \end{align}

Using  (15) and (16) in (14) we get (for the details see  Appendix B) the equation  relating  parameters  $\mu_1,\mu_2,\mu_3,\mu_4$ :
\begin{align}
\label{17}
 &\mu _2 ^2 (1 - v^2 /c^2 )( 1 - \frac{c\cos \Theta }{v})^2  + 2\frac{\mu _2 }{\mu _1 }\frac{c\cos \Theta }{v}( 1 - \frac{c\cos \Theta }{v}) +  \nonumber\\
  &+ ( \frac{c\cos \Theta }{\mu _1 v})^2  - (\alpha ^2  + \beta ^2 )\sin ^2 \Theta  = 0
 \end{align}
Solution of this quadratic equation for $\mu _2  = f(\mu _1 ,\mu _3 ,\mu _4 ,\cos \Theta )$  is
	\begin{align}
\label{18}
 \mu _2 ^ +   &=   \frac{\gamma ^2 }{\mu _1 }\frac{(v/c)\sqrt {\cos ^2 \Theta  + (\mu _1 /\gamma )^2 (\alpha ^2  + \beta ^2 )\sin ^2 \Theta } - \cos \Theta }{(v/c) - \cos \Theta }, \nonumber\\
 &\\
 \mu _2 ^ -   &=  - \frac{\gamma ^2 }{\mu _1 }\frac{(v/c)\sqrt {\cos ^2 \Theta  + (\mu _1 /\gamma )^2 (\alpha ^2  + \beta ^2 )\sin ^2 \Theta }  + \cos \Theta }{(v/c)- \cos \Theta } \nonumber
 \end{align}
where we use
\begin{align*}
\gamma  \equiv \frac{1}{{\sqrt {1 - v^2 /c^2 } }}
\end{align*}
	 	
The choice of the physically meaningful solution in (\ref{18}) is made by considering the $1$-dimensional limit , (equivalent to $\Theta=0$ ,  that is $\cos\Theta=1$)  of  $\mu_2^+$  and $\mu_2^-$  .  As a result we obtain
\begin{align}
\label{19}
 \mu _2 ^ +  & = \mathop {Lim}\limits_{\theta  \to 0} \mu _2 ^ +   = \frac{{\gamma ^2 }}{{\mu _1 }}\frac{{v/c - 1}}{{v/c - 1}} \to \mu _1 \mu _2 ^ +   = \gamma ^2 \nonumber\\
&\\
 \mu _2 ^ - &  = \mathop {Lim}\limits_{\theta  \to 0} \mu _2 ^ -   =  - \frac{{\gamma ^2 }}{{\mu _1 }}\frac{{v/c + 1}}{{v/c - 1}} \to \mu _1 \mu _2 ^ +   = \gamma ^2 \frac{{1 + v/c}}{{1 - v/c}} = \frac{1}{{(1 - v/c)^2 }} \nonumber
 \end{align}
As was shown in [1] the one-dimensional relation the between  $\mu_1$  and $\mu_2$  is $\mu_1\mu_2=\gamma^2$ . Hence $\mu_2^+$  is the physical solution of (\ref{18}), while $\mu_2^-$   must be discarded as a spurious root.

There are $4$ parameters defining transformations (9), (11) - (13), namely  $\mu_1,\mu_2,\mu_3,\mu_4$ . Since these parameters must be independent of the velocities ${\bf{u}}$  and  ${\bf{u'}}$ ( we consider a linear transformation) , therefore these parameters must be independent of the variable angle  $\Theta$ between {\bf{r}} and {\bf{v}}. From the expression (18) for  $\mu_2^+\equiv \mu_2$ follows that this is possible if and only if
\begin{equation}
	 \label{20}
\mu_1^2(\alpha^2+\beta^2)=\gamma^2
\end{equation}
which implies
\begin{equation}
\label{21}
\mu_1\mu_2=\gamma^2
\end{equation}
exactly like in the $1$-dimensional case [1].

Thus we have obtained $2 $ conditions (20), (21) for $4 $ parameters. Since all the parameters, entering the transformations (9), (11), (12) and (13) are  constant dimensionless scalar quantities ,they can depend, at the most, only on the ratio of two constant parameters $v$ and $c $, that is $v/c$.  However, motion in the direction perpendicular to the vector {\bf{v}} is independent of this vector. Therefore constant scalar dimensionless parameters   $\mu_3$  and $\mu_4$ , characterizing motion perpendicular to {\bf{v}} , cannot contain $ v$, and therefore must be  pure numbers independent of both $v$ and $c$.

Hence the transition to the classical case (where there is no upper bound on the speed of a particle, that is $c\to\infty$ ) will keep these parameters unchanged resulting in the classical Galileo transformation, where  .This means that  in (\ref{20}) parameters $\alpha=1,\beta=0$. Now, with the help of  (\ref{20}), (\ref{21}), we get  the values of all the parameters, entering  the transformations  (\ref{9}), (\ref{11}) - (\ref{13})
\begin{equation}
\label{22}
\mu_1=\mu_2\equiv\mu=\gamma, \hspace{10pt}\mu_3=1,\mu_4=0
\end{equation}
	 	
Inserting  (22) in (9), (11) - (13) we obtain  the familiar 3-D Lorentz transformations
\begin{equation}
\label{23}
 {\bf r}(t)= {\bf r}'(t') + {\bf v}[(\gamma  - 1)\frac{{{\bf r}'(t')\bullet{\bf v}}}{{v^2 }} + \gamma t']
\end{equation}
\begin{equation}
\label{24}
 t = \gamma [t' + \frac{{{\bf r}'(t')\bullet{\bf v}}}{{c^2 }}]
 \end{equation}
and
\begin{equation}
\label{25}
 {\bf r'}(t') = {\bf r}(t) + {\bf v}[(\gamma  - 1)\frac{{{\bf r}(t)\bullet{\bf v}}}{{v^2 }} - \gamma t]
\end{equation}
\begin{equation}
\label{26}
 t '= \gamma [t- \frac{{{\bf r}(t)\bullet{\bf v}}}{{c^2 }}]
 \end{equation}
Equations  (23) and  (25) can be rewritten as follows
\begin{equation}
\label{27}
\ {\bf r}(t) = \gamma [{\bf r}'(t') - \frac{\gamma }{{\gamma  + 1}}\frac{{{\bf v} \times ({\bf r}'(t') \times {\bf v})}}{{c^2 }} + {\bf v}t']
\end{equation}
\begin{equation}
\label{28}
\ {\bf r'}(t') = \gamma [{\bf r}(t) - \frac{\gamma }{{\gamma  + 1}}\frac{{{\bf v} \times ({\bf r}(t) \times {\bf v})}}{{c^2 }} -{\bf v}t]
\end{equation}
By differentiating , (\ref{24})  and  (27), and taking into account $d{\bf{r}}(t)/dt={\bf{u}}(t)$,  $d{\bf{r'}}(t')/dt'={\bf{u'}}(t')$, we arrive at   the general velocity composition theorem for arbitrary  velocity vectors ${\bf{u}},{\bf{u'}}$ (not necessarily parallel to the translational velocity ${\bf{v}}$). After some algebra we obtain
	\begin{equation}
\label{29}
{\bf u}(t) = \frac{{{\bf u}'(t') + {\bf v}}}{{1 + {\bf v\bullet u}'(t')/c^2 }} - \frac{{\gamma /(\gamma  + 1)}}{{c^2 }}\frac{{{\bf v} \times [{\bf u}'(t')}\times{\bf v} ] }{{1 + {\bf v\bullet u}'(t')/c^2 }}
\end{equation}

The appearance (in \ref{29}) of the term containing rotations is very suggestive  indicating  the kinematic character of the Thomas precession \cite{4}. To see that let us write this term with accuracy to $1/c^2$
\begin{equation}
\thicksim\frac{1}{2 }\frac{{\bf v} \times [{\bf u}'(t')\times{\bf v} ] }{c^2 }\nonumber
\end{equation}
It is seen that  point $D$ moving with velocity {\bf{u'}} with respect to point $B$ ( in our physical setup of the compound motion of $D$)  tends to rotate towards the direction of the axis along  {\bf{v}}=$const $ and precesses  about this axis. The rate of the precession is then (with the same accuracy)
\begin{equation}
\thicksim\frac{1}{2 }\frac{(d{\bf u}'(t')/dt')\times{\bf v}}  {c^2 }\nonumber
\end{equation}

As a next step we find the invariant of the Lorentz transformation (23)-(26). To this end we use (\ref{24}) ( written in the differential form) and the expression (29) for ${\bf{u}}^2$ . With the help of these two relations we  eliminate from (\ref{24}) quantity {\bf{v}}. From (29) follows
\begin{equation}
\label{30}
1-\frac{{\bf u}^2 }{c^2} = (1-v^2/c^2)\frac{1-u'^2/c^2}{(1 + {\bf {u}}' \bullet {\bf {v}}/c^2 )^2 }, \hspace {10pt} v^2\equiv{\bf v}^2, \hspace{5pt}u'^2\equiv{\bf {u'}}^2
\end{equation}
On the other hand, from (24) we find
\begin{equation}
\label{31}
\frac{dt'^2 }{dt^2 } = \frac{1 - v^2 /c^2 }{(1 + {\bf {u}}' \bullet {\bf{ v}}/c^2 )^2 },  \hspace{5pt} v^2 \equiv {\bf {v}}^2
\end{equation}
Finally, inserting (\ref{30}) in (\ref{31}), we obtain the invariant of the Lorentz transformation (\ref{23})- (\ref{26})
\begin{equation}
\label{32}
dt^2 (1 - u^2 /c^2 ) = dt'^2 (1 - u'^2 /c^2 )
\end{equation}

This can be viewed as the Poincare's  "complicated rule" accounting for the difference between $t$  and $t'$  that  was conjectured ( but not found) by him [3], and  which we derived  using classical kinematics , as a consequence of the existence of the finite upper limit of a particle's speed $ c$. It has turned out  that the existence of the  invariant (\ref{32}) ( not present in classical mechanics where there is no upper bound on a speed of a particle) is of crucial importance for the dynamics of a particle with the upper bound on its speed.

\section{Conclusion}

 \hspace{20pt}An introduction into classical kinematics of a single constraint, imposed on the magnitude of a speed of a particle, resulted in a revision of the theorem  of vectors addition, prescribed by classical kinematics. A critical look at  this theorem  reveals that it is blindly borrowed from classical geometry of objects at rest.  Its modification, which takes  into account the fact that in this new kinematics the objects are not at rest anymore (and, in addition their speed is bounded from above) led to the derivation of the well-known  relations of the special relativity( the Lorentz transformations and the velocity composition theorem). on the surface this  seems like one more only addition to a host of such derivations. However our derivation not only methodologically but $conceptually$ differs  from the rest of the derivations  universally  derived with the help of the basic concepts of the special relativity. By contrast, we use the conventional classical kinematics ( albeit modified by the crucial assumption of the existence of the upper limit on the speed of a particle, $c$) without making an  $apriori $ assumptions about time dilatation in  uniformly moving frames of references. Such  an approach makes the Lorentz transformations and its respective corollaries an extension of classical kinematics into the realm of kinematics( still using classical time and space), by assuming the existence of the upper bound on a particle's speed , not present  in classical kinematics and employing the major concept of  a geometry of moving particles, ${AD\neq AB+BD}$. These two concepts lead to a drastic change of  the conventional concepts of  classical kinematics,  such as, for example ,the  vector addition.

\appendix
\numberwithin{equation}{section}
\section{Appendix A}
Derivation of (\ref{13})

By constructing the cross product of (\ref{9}) and {\bf v} we get
\begin{equation}
 \label{A1}
 {\bf r} \times {\bf v} = \mu _4 ({\bf v}\frac{{{\bf r}'\bullet{\bf v}}}{v} - {\bf r}'v) + \mu _3 v({\bf r}' \times {\bf v})
 \end{equation}
 As the next step we find ${\bf {r}}' \bullet{\bf v}$. From the same relation (9) we find
 	  \begin{equation}
 \label{A2}
 {\bf r}\bullet {\bf v} = \mu _1 ({\bf r}'\bullet{\bf v} + v^2 t')
 \end{equation}
Hence
\begin{equation}
\label{A3}
{\bf r}'\bullet{\bf v} = \frac{{{\bf r}\bullet{\bf v}}}{{\mu _1 }} - v^2 t'
\end{equation}
Inserting in (\ref{A3}) the expression for  $t'$  from (\ref{11}) we obtain
\begin{equation}
\label{A4}
{\bf r}'\bullet{\bf v} = \mu _2 ({\bf r}\bullet{\bf v} - v^2 t)
\end{equation}
By using (\ref{A4}) in (\ref{A1}) we find
\begin{equation}
\label{A5}
{\bf r}' \times {\bf v} = \frac{{\mu _4 }}{{\mu _3 }}{\bf r} \times {\bf v} + [{\bf r}'v - \mu _2 \frac{{\bf v}}{v}({\bf r\bullet v} - v^2 t)]
\end{equation}
Inserting (A4) and (A5) back in (9) we arrive (after rather  lengthy, but straightforward calculations) at the following
 	\begin{equation}
 \label{A6}
 {\bf r} = \frac{{\mu _3 ^2  + \mu _4 ^2 }}{{\mu _3 }}{\bf r}' + \frac{{\mu _4 }}{{\mu _3 }}\frac{{{\bf r} \times {\bf v}}}{v} + \frac{{\bf v}}{{v^2 }}{\bf r} \bullet {\bf v}(1 - \mu _2 \frac{{\mu _3 ^2  + \mu _4 ^2 }}{{\mu _3 }}) + \mu _2 \frac{{\mu _3 ^2  + \mu _4 ^2 }}{{\mu _3 }}{\bf v}t
 \end{equation}
From (A6) immediately follows expression (13)

\renewcommand{\thesection}{B}

\section{Appendix B}	
Derivation of (\ref{17})

\begin{align}
\label{B1}
( \frac{r'^2 }{t'^2 })_{max} & = \frac{\mu _2 ^2 (r\cos \Theta  - vt)^2  + r^2 \sin ^2 \Theta /\mu _3 ^2 }
{ \mu _2 ^2 t^2  + ( [1 - \mu _1 \mu _2 ]/\mu _1  )^2 ( r\cos \Theta/v)^2  + 2\mu _2 t r\cos \Theta(1 - \mu _1 \mu _2 )/\mu _1  v} \equiv\nonumber\\
   &\equiv\frac{\mu _2 ^2 (c\cos \Theta  - v)^2  + c^2 \sin ^2 \Theta /\mu _3 ^2 }
 {\mu _2 ^2  + ([1 - \mu _1 \mu _2 ]/\mu _1  )^2 ( c\cos \Theta/v) ^2  + 2\mu _2 c\cos \Theta(1 - \mu _1 \mu _2 )/{\mu _1 v}} \equiv \nonumber\\
  &\equiv c^2\frac{\mu _2 ^2 (v/c)^2 (c\cos \Theta /v - 1)^2  + \sin ^2 \Theta \mu _3 ^2 }
{ \mu _2 ^2  +  ( [1 - \mu _1 \mu _2]/ \mu _1  )^2 (c\cos \Theta /v )^2  + 2\mu _2 c\cos \Theta(1 - \mu _1 \mu _2 )/\mu _1v}= c^2
\end{align}
We identically rewrite the denominator of (B1):

\begin{align}
\label{B2}
&\mu _2 ^2  + ( {\frac{{1 - \mu _1 \mu _2 }}{{\mu _1 }}} )^2 ( {\frac{{c\cos \Theta }}{v}} )^2  + 2\mu _2 \frac{{1 - \mu _1 \mu _2 }}{{\mu _1 }}\frac{{c\cos \Theta }}{v}\equiv\nonumber\\
&\equiv\mu _2 ^2  + ( {\mu _2 ^2  - 2\frac{{\mu _2 }}{{\mu _1 }} + \frac{1}{{\mu _1 ^2 }}} )( {\frac{{c\cos \Theta }}{v}})^2  + 2\frac{{\mu _2 }}{{\mu _1 }}\frac{{c\cos \Theta }}{v} - 2\mu _2 ^2 \frac{{c\cos \Theta }}{v}+\nonumber\\
&+ ( {\frac{{c\cos \Theta }}{{\mu _1 v}}})^2  \equiv \mu _2 ^2 t( {1 - \frac{{c\cos \Theta }}{v}} )^2  + 2\frac{{\mu _2 }}{{\mu _1 }}\frac{{c\cos \Theta }}{v}( {1 - \frac{{c\cos \Theta }}{U}}) + ( {\frac{{c\cos \Theta }}{{\mu _1 v}}})^2
\end{align}
Inserting (\ref{B2})in (\ref{B1})we get
\begin{align}
\label{B3}
&\mu _2 ^2 (\frac{U}{c})^2 (\frac{{c\cos \Theta }}{v} - 1)^2  + \frac{{\sin ^2 \Theta }}{{\mu _3 ^2 }} = \mu _2 ^2 \left( {1 - \frac{{c\cos \Theta }}{v}} \right)^2+\nonumber\\
 &+ 2\frac{{\mu _2 }}{{\mu _1 }}\frac{{c\cos \Theta }}{U}\left( {1 - \frac{{c\cos \Theta }}{U}} \right) + \left( {\frac{{c\cos \Theta }}{{\mu _1 U}}} \right)^2
\end{align}
By gathering the similar terms we obtain equation (\ref{17})

\end{document}